\documentclass{article}
\usepackage{cite}
\usepackage{spconf,amsmath,graphicx,hyperref}
\usepackage{graphicx} % Required for inserting images
\usepackage{siunitx}
\usepackage{subcaption}
\usepackage{multirow}
\usepackage{booktabs}
\usepackage{url}
\usepackage{bm}
\usepackage{cleveref}
\usepackage{amsmath} % Required for some math elements 
\usepackage{amsfonts}
\usepackage{comment}
\newcommand{\txin}{\text{(in)}}
\newcommand{\txout}{\text{(out)}}
\newcommand{\Fs}{F_\text{s}}
\newcommand{\sinc}[0]{\text{sinc}}
\newcommand{\ktrainable}[0]{k_\text{(tr)}}
\newcommand{\kwinsinc}[0]{k_\text{(win-sinc)}}
\newcommand{\knoisy}[0]{k_\text{(noisy)}}
\newcommand{\ldnn}[0]{\mathcal{L}}

\title{Dissecting Performance Degradation in Audio Source Separation \\under Sampling Frequency Mismatch}
\name{Kanami Imamura$^{1,2}$, Tomohiko Nakamura$^2$, Kohei Yatabe$^3$, and Hiroshi Saruwatari$^1$\thanks{This work was supported by JSPS KAKENHI under Grant JP23K28108 and JST PRESTO under Grant number JPMJPR2517.}}
\address{$^1$ The University of Tokyo, Japan\\
$^2$ The National Institute of Advanced Industrial Science and Technology (AIST), Japan \\
$^3$ Tokyo University of Agriculture and Technology, Japan
}

\begin{document}
\ninept
\maketitle
\begin{abstract}
Audio processing methods based on deep neural networks are typically trained at a single sampling frequency (SF).
To handle untrained SFs, signal resampling of the input signal is commonly employed, but it can degrade performance, particularly when the input SF is lower than the trained SF. 
This paper investigates the causes of this degradation through two hypotheses: (i) the lack of high-frequency components introduced by up-sampling, and (ii) the greater importance of their presence than their spectral contents. 
To examine these hypotheses, we compare conventional resampling with three alternatives: post-resampling noise addition, which adds Gaussian noise to the resampled signal; noisy-kernel resampling, which perturbs the kernel with Gaussian noise to enrich high-frequency components; and trainable-kernel resampling, which adapts the interpolation kernel through training.
Experiments on music source separation show that noisy-kernel and trainable-kernel resampling alleviate the degradation observed with conventional resampling.
We further demonstrate that noisy-kernel resampling is effective across diverse models, highlighting it as a simple yet practical option to alleviate the performance degradation.
\end{abstract}
\begin{keywords}
Resampling, sampling frequency, audio source separation, deep learning
\end{keywords}
\section{Introduction} \label{sec:intro}
Audio signal processing methods based on deep neural networks (DNNs) are typically trained and evaluated at a single sampling frequency (SF).
To handle untrained SFs, the common solution is to resample the input signal to the trained SF, process it by the DNN, and resample the output back to the original SF.
Recent studies have shown that such resampling can degrade performance in DNN-based audio source separation methods, particularly when the input SF is lower than the trained SF~\cite{saito2022sficonvtasnet,imamura2023noninteger,yu2023sfibsrnn,zhang2024urgent}.
These studies mainly focus on specific models and conditions.
Thus, a comprehensive analysis is missing.
As a result, the mechanism behind this degradation remains unclear, which motivates a systematic study of its causes and remedies.

To address this issue, we consider two hypotheses about the cause of the performance degradation.
The first hypothesis is that \emph{the lack of high-frequency components caused by up-sampling leads to performance degradation.}
When the input SF is lower than the trained SF, up-sampling produces signals whose frequency components above the input Nyquist frequency are set to zero.
This absence of high-frequency components mismatches the training distribution and can impair performance.
The second hypothesis is that \emph{the presence of high-frequency components contributes more to separation performance than their spectral contents}.
DNN-based source separation models generally focus on low-frequency components, with coarse resolution at high frequencies.
For example, in Conv-TasNet, the encoder allocates more channels to process the low-frequency components~\cite{luo2019convtasnet}. 
Similar trends appear in its extensions~\cite{ditter2020mpgtf,saito2022sficonvtasnet,imamura2024naf}.
Furthermore, the band-split recurrent neural network (BSRNN)~\cite{luo2023bsrnn} explicitly assigns finer resolution to low-frequency bands and coarser resolution to high-frequency bands.
These models potentially place less emphasis on high-frequency components.
Despite the necessity of their presence, their spectral contents may be less critical.

To examine these hypotheses, we introduce three resampling methods and compare them with the conventional resampling.
The first method adds random noise directly to the resampled signal to supplement missing high-frequency components.
The second method perturbs the interpolation kernel with random noise, which enriches the high-frequency components of the resampled signal.
The third method parameterizes the interpolation kernel with a DNN.
While the original separation network is kept frozen, only the network representing the kernel is trained so that the resampling process adapts to the model and mitigates performance degradation.
Through music source separation experiments, we demonstrate that our two hypotheses are supported and that noisy-kernel resampling consistently mitigates degradation across diverse models.

\begin{figure*}[t]
    \centering
    \begin{subfigure}{0.9\linewidth}
        \centering
        \includegraphics[width=\linewidth]{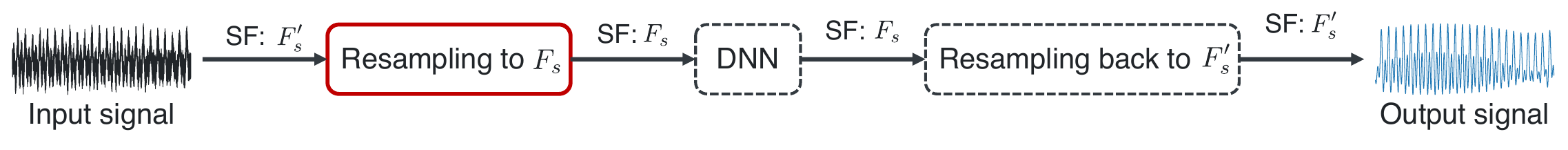}
    \vspace{-1.5em}
        
        \caption{Processing pipeline of DNN handling untrained SF with resampling.}
        \label{fig:proposed_pipeline}
    \end{subfigure}
    \\
    \begin{subfigure}{0.31\linewidth}
        \centering
        \includegraphics[width=\linewidth]{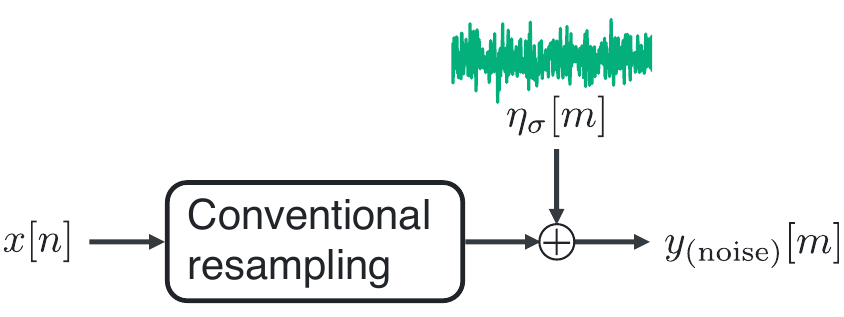}
        \caption{Post-resampling noise addition.}
        \label{fig:proposed_noise}
    \end{subfigure}
    \hfill
    \begin{subfigure}{0.32\linewidth}
        \centering
        \includegraphics[width=\linewidth]{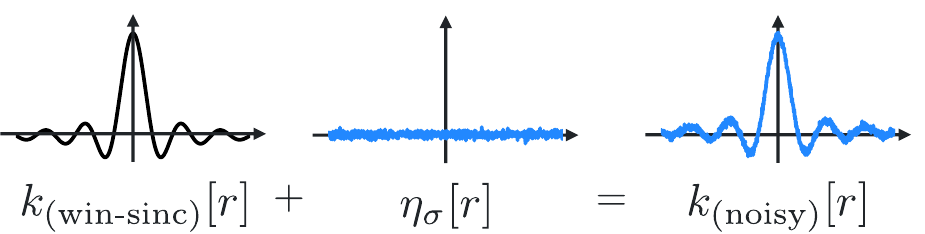}
        \caption{Noisy-kernel resampling.}
        \label{fig:proposed_noisy}
    \end{subfigure}
    \hfill
    \begin{subfigure}{0.31\linewidth}
        \centering
        \includegraphics[width=\linewidth]{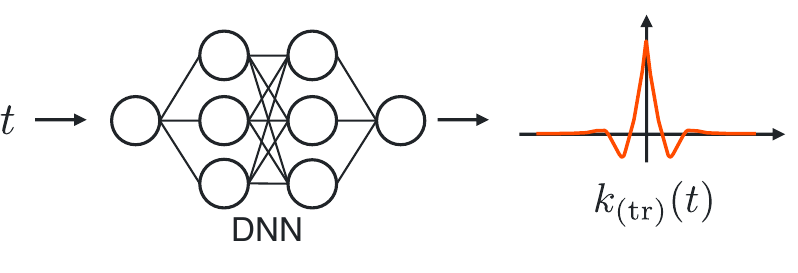}
        \caption{Trainable-kernel resampling.}
        \label{fig:proposed_trainable}
    \end{subfigure}
    \vspace{-0.3em}
    \caption{Overview of processing pipeline and kernels in proposed resampling methods.}
    \label{fig:proposed}
\end{figure*}

\section{Related Works}
\subsection{Interpolation-Based Resampling}
Interpolation-based resampling is one of the most widely used approaches.
Let $(x[n])_{n=0}^{N-1}$ denote the input signal of length $N$ sampled at the SF $\Fs^\txin$.
After resampling to the SF $\Fs^\txout$, we obtain $(y[m])_{m=0}^{M-1}$ of length $M=\lfloor \Fs^\txout N /\Fs^\txin \rfloor$, where $\lfloor\cdot\rfloor$ denotes the floor function:
\begin{equation}
    y[m] = \sum_{n=0}^{N-1} x[n] \> k\left(\dfrac{m}{\Fs^\txout} - \dfrac{n}{\Fs^\txin}; \Fs^\txin \right), \label{eq:resampling}
\end{equation}
where $k(t,\Fs)$ is the interpolation kernel.
A common choice is the windowed sinc function, 
\begin{equation}
    \kwinsinc(t;\Fs)=z(t)\sinc(\Fs t), \quad \sinc(t) = \dfrac{\sin(\pi t)}{\pi t}, \label{eq:sinc}
\end{equation}
with $t\in\mathbb{R}$ denoting continuous time and $z(t)$ a window function.
We use a non-negative smooth window $z(t)$ and assume $z(t)=0$ for $t < -L/(2\Fs)$ or $t > L/(2\Fs)$, where $L$ is a positive integer.
The Kaiser window is a typical choice.
The window length $L$ controls the trade-off between interpolation accuracy and computational cost.
Since the sinc function acts as an ideal low-pass filter with cutoff frequency $\Fs/2$, the windowed sinc kernel suppresses frequency components above the Nyquist frequency.

\subsection{Impact of Resampling on DNN-Based Methods}
As described in \Cref{sec:intro}, the performance degradation due to signal resampling has been reported. 
However, it has been examined only in a few studies.
In~\cite{saito2022sficonvtasnet}, a Conv-TasNet adaptation for music source separation was evaluated under different SFs, and degradation was observed especially when the input SF was lower than the trained SF.
In related work on speech enhancement, \cite{yu2023sfibsrnn} evaluated resampling effects on the performance of BSRNN.
In addition, \cite{zhang2024urgent} reported results across different SFs for models submitted to the speech enhancement competition.
The reported impact varied across architectures, and degradation was not always but unpredictably observed.
In~\cite{carson2025resampling}, resampling effects were also analyzed, but the scope was limited to recurrent neural networks for neural audio effect processing.

\section{Methods}
To examine the hypotheses described in \Cref{sec:intro}, we introduce three resampling methods.
\Cref{fig:proposed} shows an overview of the methods.
We denote the input and trained sampling frequencies by $\Fs'$ and $\Fs$, respectively, and focus on the case $\Fs' < \Fs$, where degradation due to resampling has been observed.

\begin{figure*}[t]
    \centering
    \includegraphics[width=0.9\linewidth]{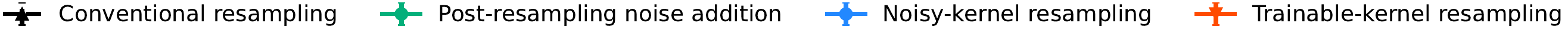}
    \\
    \vspace{0.4em}
    \centering
         \begin{minipage}{0.242\hsize}
            \centering
            \includegraphics{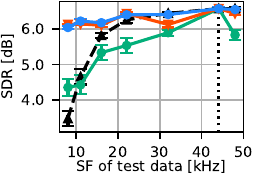}
            \subcaption{vocals}
        \end{minipage}
        \begin{minipage}{0.242\hsize}
            \centering
            \includegraphics{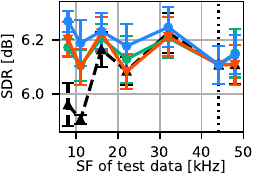}
            \subcaption{bass}
        \end{minipage}
        \begin{minipage}{0.242\hsize}
            \centering
            \includegraphics{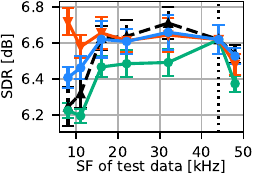}
            \subcaption{drums}
        \end{minipage}
        \begin{minipage}{0.242\hsize}
            \centering
            \includegraphics{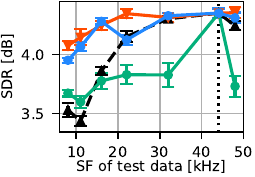}
            \subcaption{other}
        \end{minipage}
    \vspace{-0.3em}
    \caption{
        SDRs of conventional resampling, post-resampling noise addition, noisy-kernel resampling, and trainable-kernel resampling.
    }
    \label{fig:sdr-resampling}
\end{figure*}

\subsection{Post-Resampling Noise Addition} \label{sec:resampling_plus_noise}
The first method is a simple baseline that directly adds Gaussian noise to the signal obtained after conventional resampling. 
The output signal is given as
\begin{equation}
    y_\text{(noise)}[m] = \sum_{n=0}^N x[n] \> \kwinsinc\left(\dfrac{m}{\Fs} - \dfrac{n}{\Fs'}; \Fs' \right) + \eta_\sigma[m],
\end{equation}
where $\eta_\sigma[m]$ is zero-mean Gaussian noise with variance $\sigma^2$, sampled independently at each time index.
This operation injects energy across the entire frequency range and supplements the components above the Nyquist frequency.

\subsection{Noisy-Kernel Resampling} \label{sec:noisy_kernel}
The second method modifies the conventional windowed-sinc interpolation kernel by adding Gaussian noise.
In this method, Gaussian noise is added to a discretized kernel in the interpolation computation.
Let $r$ denote the discrete-time index of the kernel.
The kernel $\knoisy[r;\Fs',\sigma]$ is defined as
\begin{equation}
    \knoisy[r;\Fs',\sigma] = \kwinsinc[r;\Fs'] + \eta_\sigma[r].
\end{equation}
The frequency response of $\knoisy[r;\Fs',\sigma]$ is the sum of the responses of $\kwinsinc[r;\Fs']$ and $\eta_\sigma[r]$.
Consequently, compared with the conventional kernel, $\knoisy[r;\Fs',\sigma]$ shows a larger magnitude in the frequency band above the cutoff frequency $\Fs'/2$.
The strength of these high-frequency components depends on $\sigma$.

\subsection{Trainable-Kernel Resampling} \label{sec:trainable_kernel}
The third method is the only data-driven approach considered in this study.
It represents the interpolation kernel as a continuous-time function using a DNN.
Given time $t$ as input, the network $\ktrainable(t;\theta):\mathbb{R}\rightarrow\mathbb{R}$ outputs the kernel amplitude, where $\theta$ denotes the network parameters.
Such implicit function representations have also been shown to be effective in image and audio processing~\cite{kim2022sr,imamura2024naf,sitzmann2020inrsiren,mildenhall2021nerf}, which motivates us to use this formulation.

During training, only the kernel network is updated, while the separation model, pretrained at $\Fs$, remains frozen.
For simplicity, we adopt a multilayer perceptron (MLP) as the kernel network.
The parameters $\theta$ can be optimized by backpropagation.
For a differentiable loss function $\ldnn$, the gradient with respect to 
$\theta$ can be analytically derived, demonstrating that $\ktrainable(t;\theta)$ can be updated in an end-to-end manner as follows:
\begin{align}
    \dfrac{\partial \ldnn}{\partial\theta} 
    &= \sum_{m=0}^M \dfrac{\partial \ldnn}{\partial y[m]} \dfrac{\partial y[m]}{\partial \theta} \\
    &= \sum_{m=0}^M \dfrac{\partial \ldnn}{\partial y[m]} \sum_{n=0}^N x[n] \dfrac{\partial}{\partial \theta} \ktrainable\left(\dfrac{m}{\Fs} - \dfrac{n}{\Fs'}; \theta \right).
\end{align}

The loss function is defined as the sum of separation and resampling losses:
\begin{equation}
    \ldnn = \|\hat{\bm s}-\bm s\|^2_2 + \|\bm y_\text{(tr)}-\bm y_\text{(win-sinc)}\|^2_2.
\end{equation}
The first term measures the separation errors between the ground-truth source signals $\bm{s}$ and the estimated sources $\bm{{\hat{s}}}$.
The estimated sources $\bm{{\hat{s}}}$ are obtained by resampling the input signal to $\Fs$ with $\ktrainable(t;\theta)$, processing it with the separation model, and resampling the output back to $\Fs'$.
The second term regularizes the resampling process by minimizing the difference between $\bm y_\text{(tr)}$, the signal resampled with $\ktrainable(t;\theta)$, and $\bm y_\text{(win-sinc)}$, the signal resampled with the conventional windowed-sinc kernel.

\section{Experimental Setup}
\subsection{Dataset}
We conducted experiments on music source separation using the MUSDB18-HQ dataset~\cite{rafii2019musdbhq}, a widely used benchmark in this field.
It consists of 150 stereo tracks with four sources (vocals, bass, drums, and other).
We followed the official data split, using 86 tracks for training, 14 for validation, and 50 for testing.

During training and validation, all tracks were used at the original SF of \SI{44.1}{\kilo\hertz}.
For testing, in addition to \SI{44.1}{\kilo\hertz}, we evaluated resampled versions of the test set at 8, 11.025, 16, 22.05, and \SI{32}{\kilo\hertz} to examine the effect of SF mismatch.

\begin{figure}[t]
    \centering
    \includegraphics[width=0.8\linewidth]{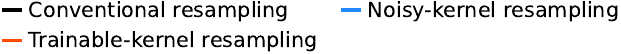}
    \\
    \vspace{0.4em}
    \centering
    \begin{minipage}{0.48\hsize}
        \centering
        \includegraphics{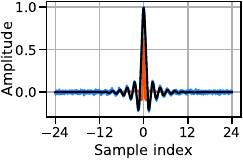}
        \subcaption{Waveform}
        \label{fig:kernel_time}
    \end{minipage}
    \hspace{0.3em}
    \begin{minipage}{0.48\hsize}
        \centering
        \includegraphics{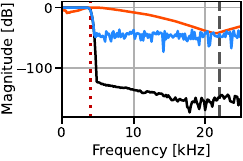}
        \subcaption{Frequency response}
        \label{fig:kernel_freq}
    \end{minipage}
    \vspace{-0.3em}
    \caption{Comparison of kernel functions of conventional resampling, noisy-kernel resampling, and trainable-kernel resampling for vocals for 8~kHz.
    Red dotted line and gray dashed line denote the input and trained Nyquist frequencies, respectively.
    }
    \label{fig:kernel}
\end{figure}
\begin{figure}[t]
    \centering
    \begin{subfigure}{0.48\linewidth}
        \centering
        \includegraphics[width=\linewidth]{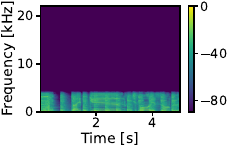}
        \vspace{-1em}
        \caption{Conventional resampling.}
        \label{fig:spec_conventional}
    \end{subfigure}
    \hspace{0.02\linewidth}
    \begin{subfigure}{0.48\linewidth}
        \centering
        \includegraphics[width=\linewidth]{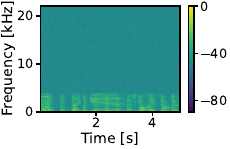}
        \vspace{-1em}
        \caption{Post-resampling noise addition.}
        \label{fig:spec_noise}
    \end{subfigure}
    \\
    \begin{subfigure}{0.48\linewidth}
        \centering
        \includegraphics[width=\linewidth]{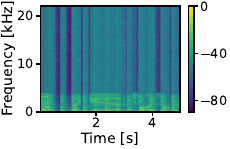}
        \caption{Noisy-kernel resampling.}
        \label{fig:spec_noisy_kernel}
    \end{subfigure}
    \hspace{0.02\linewidth}
    \begin{subfigure}{0.48\linewidth}
        \centering
        \includegraphics[width=\linewidth]{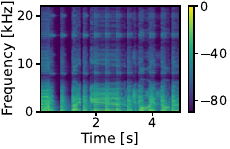}
        \caption{Trainable-kernel resampling.}
        \label{fig:spec_trainable}
    \end{subfigure}
    \vspace{-0.3em}
    \caption{Spectrogram examples of signals resampled from \SI{8}{\kilo\hertz} to \SI{44.1}{\kilo\hertz} with the compared methods.}
    \label{fig:spectrogram}
\end{figure}

\begin{table*}[t]
    \centering
    {\footnotesize
    \caption{SDRs~[\si{\decibel}] of various models at \SI{44.1}{\kilo\hertz}, at \SI{8}{\kilo\hertz} with conventional resampling, and at \SI{8}{\kilo\hertz} with noisy-kernel resampling. 
    SDRs at \SI{8}{\kilo\hertz} with conventional resampling that are lower than those at \SI{44.1}{\kilo\hertz} are underlined, and SDRs at \SI{8}{\kilo\hertz} with noisy-kernel resampling that are higher than those with conventional resampling are shown in bold.}
    \begin{tabular}{ccccccccccccc}
        \toprule
        \multirow{2}[1]{*}{Model} & \multicolumn{4}{c}{44.1~\si{\kilo\hertz} (Trained SF)} & \multicolumn{4}{c}{8~\si{\kilo\hertz} with conventional resampling} & \multicolumn{4}{c}{8~\si{\kilo\hertz} with noisy-kernel resampling} \\
        \cmidrule(l){2-5} \cmidrule(l){6-9} \cmidrule(l){10-13}
         & vocals & bass & drums & other &  vocals & bass & drums & other &  vocals & bass & drums & other \\
        \midrule
        Conv-TasNet~\cite{saito2022sficonvtasnet} & 5.95 & 5.31 & 5.64 & 3.81 & \underline{0.91} & \underline{3.74} & \underline{3.18} & \underline{1.00} & \textbf{5.06} & \textbf{5.20} & \textbf{5.20} & \textbf{3.42}  \\        BSRNN~\cite{luo2023bsrnn} & 6.58 & 6.11 & 6.62 & 4.35 & \underline{3.47} & 5.96 & 6.25 & \underline{3.53} & \textbf{6.05} & 6.27 & 6.41 & \textbf{4.06} \\
        Mel-RoFormer~\cite{wang2024melroformer} & 9.67 & 7.01 & 8.58 & 6.67 & \underline{8.42} & \underline{6.18} & \underline{4.96} & \underline{5.62} & \textbf{9.00} & \textbf{7.24} & \textbf{6.93} & \textbf{6.13} \\
        HT-Demucs~\cite{rouard2023htdemucs} & 8.85 & 10.03 & 9.94 & 6.65 & \underline{7.98} & 9.93 & \underline{8.90} & \underline{5.74} & 8.03 & 9.95 & 9.04 & 5.95 \\
        \midrule
        BS-RoFormer~\cite{lu2024bsroformer} & 10.83 & 9.47 & 11.56 & 7.81 & 10.59 & 9.30 & \underline{10.79} & 7.22 & 10.51 & 9.41 & 10.65 & 7.37 \\
        MDX23C\footnotemark & 9.17 & 6.64 & 7.66 & 6.16 & 9.28 & 6.25 & 6.80 & \underline{5.63} & 9.26 & 6.44 & 7.03 & 5.63 \\
        SCNet~\cite{tong2024scnet} & 9.46 & 9.43 & 10.21 & 7.07 & 9.33 & 9.50 & 10.07 & 6.87 & 9.32 & 9.49 & 10.09 & 6.87 \\
        \bottomrule
    \end{tabular}
    \label{tab:compare_models}
    }
\end{table*}
\footnotetext{\url{https://github.com/kuielab/sdx23/}.}

\subsection{Implementation}
We compared four resampling methods.

\noindent
\textbf{Conventional resampling.} This method was implemented using the \path{torchaudio} library~\cite{yang2022torchaudio}.
The interpolation kernel was a windowed-sinc function with a Kaiser window.
The window length was set to $L=48$ to ensure sufficient approximation accuracy.
The Kaiser window parameter $\alpha$ was kept at the default value in \path{torchaudio.functional.resample} ($\alpha \approx 4.1$).

\noindent \textbf{Post-resampling noise addition. (\Cref{sec:resampling_plus_noise})} Gaussian noise was added after conventional resampling.
We used the same settings as conventional resampling.
The noise variance was determined so that the signal-to-noise ratio (SNR) of each resampled signal was kept constant.
The SNR was set to \SI{20}{\decibel}.

\noindent \textbf{Noisy-kernel resampling. (\Cref{sec:noisy_kernel})} Gaussian noise was added to the windowed-sinc kernel.
The noise variance was set to $\sigma^2=1.0\times 10^{-6}$, which gave the best performance in preliminary experiments.
For each input signal, noise was newly sampled to reduce the influence of randomness on the results.

\noindent \textbf{Trainable-kernel resampling. (\Cref{sec:trainable_kernel})} The kernel network was implemented as a three-layer MLP with 32 hidden units, interleaved with layer normalization and rectified linear unit activations.
It was trained with the Adam optimizer with an initial learning rate of $1.0\times10^{-3}$, decayed by 0.98 every two epochs.
Gradient clipping was applied with a maximum norm of 5.
Early stopping was applied when the best validation performance was not observed for ten consecutive epochs.
The maximum number of training epochs was set to 100 and the batch size was set to 4.

\section{Analysis} \label{sec:analysis}
\subsection{Comparison of Resampling Methods} \label{sec:compare_resampling}
In this section, we compared the four resampling methods using a recent baseline separation network, BSRNN.
The implementation followed the publicly available code\footnote{\url{https://github.com/amanteur/BandSplitRNN-PyTorch}} with the same configuration as in~\cite{luo2023bsrnn}.
To reduce initialization dependency, we trained four models with different random seeds.
We then computed the averages and standard errors of the source-to-distortion ratios (SDRs) using the BSSEval v4 toolkit~\cite{stoter2018bsseval}.

\Cref{fig:sdr-resampling} shows the SDR results for all sources.
With conventional resampling, the performance degraded most severely for vocals, where the SDR at \SI{8}{\kilo\hertz} was approximately \SI{3}{\decibel} lower than that at the trained SF.
Smaller but consistent drops were observed for drums and other (about \SI{0.8}{\decibel} and \SI{0.6}{\decibel}, respectively), whereas bass showed little degradation.
In contrast, with noisy- and trainable-kernel resampling, no degradation was observed for any source.
As shown in Fig.~\ref{fig:kernel}, the kernels of noisy- and trainable-kernel resampling contained greater energy above the Nyquist frequency, \SI{4}{\kilo\hertz}, than the kernel of conventional resampling.
As a result, the resampled signals of noisy-kernel resampling and trainable-kernel resampling had components above \SI{4}{\kilo\hertz}, unlike the resampled signal of conventional resampling.
These results support our first hypothesis.

Unlike noisy- and trainable-kernel resampling, post-resampling noise addition failed to mitigate degradation.
Its performance was comparable to, or even lower than, that of conventional resampling.
In particular, it significantly degraded for vocals and other.
As shown in Fig.~\ref{fig:spectrogram}, the signal obtained by post-resampling noise addition contained larger high-frequency components due to the added noise.
However, this noise was spread across a wide frequency range, including the band containing the original signal components, which can result in performance degradation.

With noisy-kernel resampling, the resampled signal had noise components above \SI{4}{\kilo\hertz}, correlated with the input signal components below \SI{4}{\kilo\hertz}, as shown in Fig.~\ref{fig:spectrogram}.
The resampled signal with trainable-kernel resampling showed a folded structure, where low-frequency components were aliased into the high-frequency band.
Despite these differences, noisy- and trainable-kernel resampling produced comparable separation performance.
These findings support our second hypothesis, while also suggesting that the importance of the spectral contents of high-frequency components diminishes when they are correlated with the input signal.

\subsection{Generality of Proposed Resampling Methods}
We next examined whether noisy-kernel resampling can mitigate the degradation caused by conventional resampling in other music source separation models.
We used six pretrained models published in the public music source separation repository\footnote{\url{https://github.com/ZFTurbo/Music-Source-Separation-Training}}, all designed for four-stem separation.
In addition, we tested a Conv-TasNet adaptation for music source separation\cite{saito2022sficonvtasnet}, which we refer to as \emph{Conv-TasNet} for brevity.
Since it suffers performance degradation with conventional resampling, it serves as a reference for evaluating the effectiveness of our method.
All models were trained at \SI{44.1}{\kilo\hertz}.

\Cref{tab:compare_models} shows the SDRs of the compared models at \SI{44.1}{\kilo\hertz}, at \SI{8}{\kilo\hertz} with conventional resampling, and at \SI{8}{\kilo\hertz} with noisy-kernel resampling.
The models in the upper block exhibited performance degradation greater than \SI{0.5}{\decibel} with conventional resampling for multiple instruments, while the models in the lower block did not.
Performance degradation was observed in BSRNN, Mel-RoFormer, and Conv-TasNet. 
A smaller but noticeable degradation appeared in HT-Demucs.
Noisy-kernel resampling mitigated this degradation, demonstrating its generality across separation models.
A similar trend was observed at \SI{16}{\kilo\hertz}, where conventional resampling caused smaller performance drops, and noisy-kernel resampling alleviated them.
We further observed a tendency that the models that prioritize low-frequency resolution are more likely to suffer from performance drops under conventional resampling.
This observation suggests that such architectures may be sensitive to resampling mismatch, although a detailed investigation remains as future work.

Surprisingly, when no performance degradation was observed with conventional resampling, applying noisy-kernel resampling yielded comparable results.
This indicates that the method does not harm performance in such cases and can thus be safely adopted regardless of whether degradation arises.

\section{Conclusion}
We investigated the effect of signal resampling on DNN-based music source separation under two hypotheses: (i) the lack of high-frequency components caused by up-sampling leads to performance degradation, and (ii) the presence of high-frequency components matters more than their spectral contents.
To examine these hypotheses, we introduced three resampling methods: post-resampling noise addition, which adds Gaussian noise to the resampled signal; noisy-kernel resampling, which adds Gaussian noise to a windowed sinc kernel; and trainable-kernel resampling, which models the interpolation kernel with a DNN.
Experiments showed that noisy- and trainable-kernel resampling mitigated the degradation observed with conventional resampling, supporting our hypotheses.
By contrast, the failure of post-resampling noise addition indicated the importance of the correlation between high-frequency components and the input signal.
Importantly, noisy-kernel resampling demonstrated its effectiveness across several separation models without harming performance, highlighting its practicality.

\bibliographystyle{IEEEbib}
\bibliography{abbrev,refs}

\end{document}